\documentstyle[11pt,newpasp,twoside,psfig]{article}
\def\edcomment#1{\iffalse\marginpar{\raggedright\sl#1\/}\else\relax\fi}
\reversemarginpar

\begin{document}
\title{The Evolution of $^3$He, $^4$He and D in the Galaxy}
 \author{Cristina Chiappini}
\affil{Depto. Astronomia - Observat\'orio Nacional, R. Gal. Jos\'e Cristino 77 - CEP 20921-400 - Rio de Janeiro - RJ - Brazil}
\author{Francesca Matteucci}
\affil{Dipartimento di Astronomia, Universit\`a di Trieste, Via G. B. Tiepolo 11 - 34100 Trieste, Italy}

\begin{abstract}
In this work we present the predictions of a modified version of the 
``two-infall model'' (Chiappini et al. 1997 - CMG) for the evolution of $^3$He, 
$^4$He and D in the solar vicinity, as well as their distributions
along the Galactic disk. In particular, we show that when allowing for
extra-mixing process in low mass stars (M$ <$ 2.5 $M_{\odot}$), as predicted by 
Charbonnel and do Nascimento (1998), a long standing problem in chemical evolution is
solved, namely: the overproduction of $^3$He by the chemical evolution models 
as compared to the 
observed values in the sun and in the interstellar medium.
Moreover, we show that chemical evolution models can constrain
the primordial value of the deuterium abundance and that a value of 
(D/H)$_p$ $ < $ 3 $\times$ 10$^{-5}$ is suggested by the present model. Finally,
adopting the primordial $^4$He abundance suggested by Viegas et al. (1999), we obtain
a value for $\Delta Y/\Delta Z$ $\simeq$ 2 and a better agreement with the solar
 $^4$He abundance.
\end{abstract}

\section{Introduction}
As discussed by Tosi (this volume), chemical evolution models are useful both to derive the 
primordial abundances of D, $^3$He and $^4$He and to give informations on stellar
nucleosynthesis. In this work we show the predictions of the two-infall model
(CMG) for the chemical evolution of the above elements in the solar vicinity and for 
their distribution along the galactic disk. 
We adopt a new version of the two-infall model which includes the contribution
by novae enrichment and the new proposed mechanism of extra-mixing in low mass stars (Charbonnel, 
Sackman, this meeting). The model was calibrated to the solar galactocentric distance of 8 kpc 
(we were still
adopting 10 kpc in CMG to better compare our predictions with
the ones of Matteucci and Fran\c cois 1989). This model assumes two main infall episodes for the
formation of the halo (and part of the thick disk) and thin disk, respectively. 
The timescale for the formation
of the thin disk is much longer than that of the halo, implying that the infalling
gas forming the thin disk comes not only from the halo but mainly from the intergalactic
medium. The timescale for the formation of the thin disk is assumed to be a function of the
galactocentric distance, leading to an inside-out picture for the Galaxy disk buildup.
The two-infall model differs from other models
in the literature mainly in two aspects: i) it considers an almost independent evolution
between the halo and thin disk components (see also Pagel \& Tautvaisiene 1995) and 
ii) it assumes a threshold in the star formation process (Kennicutt 1989). The last point
has important consequences for the predicted abundance gradients (Chiappini, Matteucci
\& Romano 2000 - CMR). 

\section{Results}
\subsection{The solar Vicinity}
Our present model differs from those of CMG in i) the adopted yields for the 
low and intermediate mass range stars which are now taken from van den Hoek \& Groenewegen (1997) instead
of Renzini \& Voli (1981); ii) the fact that now we are including the explosive nucleosynthesis
from nova outbursts (see Romano et al. 1999) and iii) the adopted solar galactocentric
distance of 8 kpc. Moreover, in the present model we adopt a primordial helium-4 abundance
of 0.241 (by mass) instead of 0.23 as recently suggested by Viegas et al. (1999). 

Our model is in good agreement with what is called
the minimum set of observational constraints, among which the most important is the 
G-dwarf metallicity distribution (see Tosi 2000 for a recent review; see  
Figure 1 and Tables 1 and 2).

\begin{table}
\small
\caption{Observed and predicted quantities at $R_{g,\odot}$ and $t=t_{now}$}
\begin{tabular}{llll}
\tableline\\
  & CMR & CMG & Observ. \\
\tableline\\
Metal-poor/total stars (\%)  
 & 4 \% & 6-13 \% & 2 - 10 \% \\
SNIa (century$^{-1}$) & 0.4 & 0.3 & 0.3 $\pm$ 0.2 \\
SNII (century$^{-1}$) & 1.16 & 0.78 & 1.2 $\pm$ 0.8 \\
$\Psi$(R$_{g,\odot}$,t$_{now}$) (M$_{\odot}$ pc$^{-2}$ Gyr$^{-1}$) & 2.6 & 2.6  & 2-10  \\
$\sigma_g$(R$_{g,\odot}$,t$_{now}$) (M$_{\odot}$ pc$^{-2}$) & 7.0 & 7.0 & 6.6 $\pm$ 2.5 \\
$\sigma_g$ / $\sigma_T$ (R$_{g,\odot}$,t$_{now}$) & 0.13 & 0.14 & 0.05-0.20 \\
$\dot{\sigma_{inf}}$(R$_{g,\odot}$,t$_{now}$)(M$_{\odot}$ pc$^{-2}$ Gyr$^{-1}$)  & 1.0 & 1.0 & 1.0 \\
Nova Outbursts (yr$^{-1}$) & 21 & - & 20-30 \\
X$_2(P)$/X$_2(now)$ & 1.55 & 1.5 & $<$ 3 \\
\tableline\\
\end{tabular}
\end{table}

The primordial abundances 
by mass of D and $^3$He were taken to be 4.4 $\times$ 10$^{-5}$ and 2.0 $\times$ 10$^{-5}$ respectively.
While the D primordial value is an upper limit (as can be seen in Figure 2a) the $^3$He is a lower
limit (see Figure 2b). As can be seen in Figure 2a, the observations of the local 
interstellar medium (ISM)
and the solar system represent tight constraints to the deuterium primordial abundance. In fact,
models that can reproduce the bulk of the observational data predict only a modest D destruction
(in our case a factor $\simeq$ 1.6; see Tosi et al. 1998).

\begin{table}
\small
\caption{Solar Abundances by Mass ($^*$ at 4.5 Gyrs ago)}
\begin{tabular}{llll}
\tableline\\
Element & $^*$CMG & $^*$CMR & Anders \& Grevesse (1989)\\
\tableline\\
H & 0.73 & 0.71 & 0.70 \cr
D & 4.6 (-5) & 3.3 (-5) & 4.8 (-5)  \cr
$^{3}$He & 10.0 (-5) & 2.1 (-5) & 2.9 (-5) \cr
$^{4}$He & 2.5 (-1) & 2.71 (-1) & 2.75 (-1) \cr
$^{12}$C & 1.8 (-3) & 3.5 (-3) & 3.0 (-3) \cr
$^{16}$O & 7.3 (-3) & 7.2 (-3) & 9.6 (-3) \cr
$^{14}$N & 1.4 (-3) & 1.6 (-3) & 1.1 (-3) \cr
$^{13}$C & 4.8 (-5) & 4.8 (-5) & 3.7 (-5) \cr
$^{20}$Ne & 0.9 (-3) & 1.0 (-3) & 1.6 (-3) \cr
$^{24}$Mg & 2.5 (-4) & 2.5 (-4) & 5.1 (-4) \cr
Si & 7.0 (-4) & 7.0 (-4) & 7.1 (-4) \cr
S & 3.1 (-4) & 3.0 (-4) & 4.2 (-4) \cr
Ca & 3.9 (-5) & 3.9 (-5) & 6.2 (-5) \cr
Fe & 1.43 (-3) & 1.34 (-3) & 1.27 (-3) \cr
Cu & 8.20 (-7) & 7.8 (-7) & 8.4 (-7) \cr
Zn & 2.4 (-6) & 2.3 (-6) & 2.1 (-6) \cr
Z  & 1.4 (-2) & 1.6 (-2) & 1.9 (-2) \cr
\tableline\\
\end{tabular}
\end{table}

\begin{figure}
\centerline{\psfig{figure=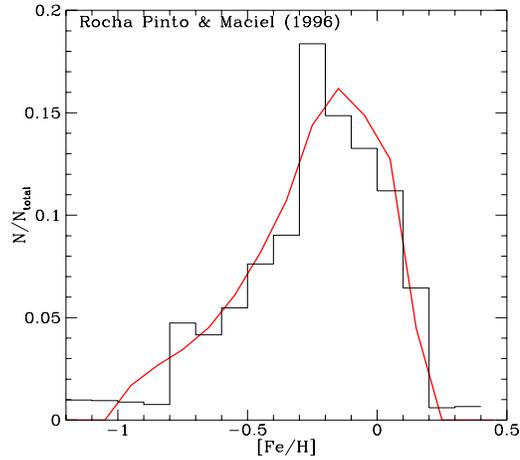,width=7.0cm,height=6.5cm}}
\caption{G-dwarf metallicity distribution at the solar vicinity. Data from Rocha-Pinto \& Maciel 1996. The
curve shows our best model prediction}
\end{figure}


With respect to our old predictions for the solar vicinity (CMG), 
we now have better agreement between the predicted solar abundances of $^3$He,
$^{12}$C and $^4$He (Table 2). The improvement in the predicted $^{12}$C is due to the fact that we 
adopt the new nucleosynthetic yields from van den Hoek \& Groenewegen (1997) for low and intermediate
mass stars. Moreover, the adoption of a higher Y primordial abundance 
(where Y is the $^4$He abundance by mass) 
leads to a better agreement with 
the solar value. The improvement in the predicted $^3$He is due to the fact that we are taking 
into account the recently proposed extra-mixing mechanism in low mass stars ($M < 2.5 M_{\odot}$, eg.
Charbonnel \& do Nascimento 1998). In Figure 2b we show models with different assumptions with 
respect to the extra-mixing mechanism. The model without extra-mixing clearly
overproduces $^3$He with respect to both solar and ISM observed abundances 
(even with the lower limit primordial abundance of $^3$He adopted in the 
present model). The two other curves show our predictions for models which 
assume that 93\% (solid line) and  99\% (dashed-line) of stars with M$<$ 2.5 M$_{\odot}$ 
completely destroy their $^3$He, respectively. A model assuming that, when a star suffers
extra-mixing (93\%) a fraction of $\simeq$ 1/100 of its $^3$He is preserved, gives essentially the same
result as the solid line model shown in Figure 2b. 
The predicted Y vs Z and Y vs O/H are shown in Figures 3a and 3b respectively. 

\begin{figure*}
\centerline{\psfig{figure=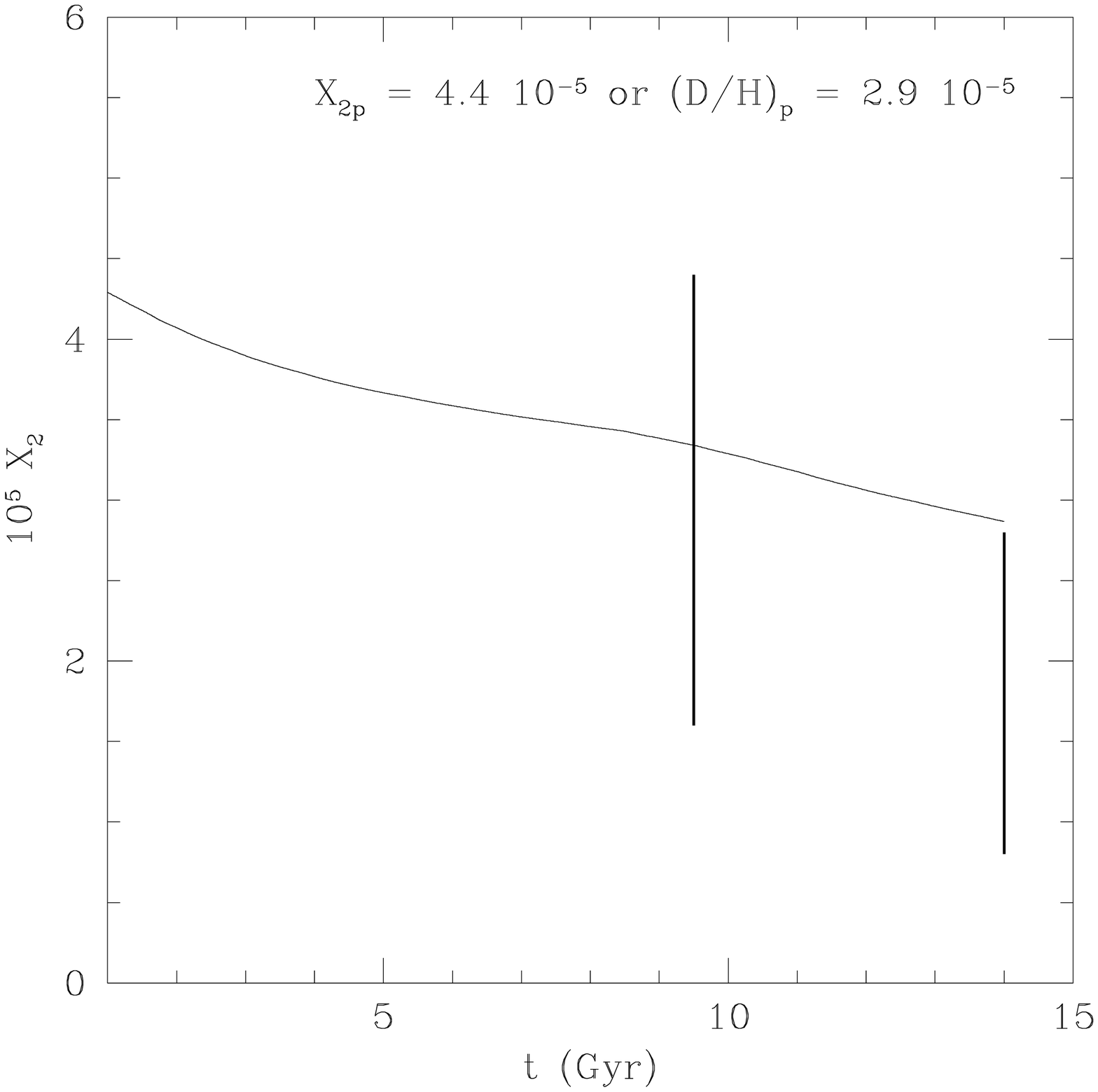,width=6.5cm,height=6cm}\psfig{figure=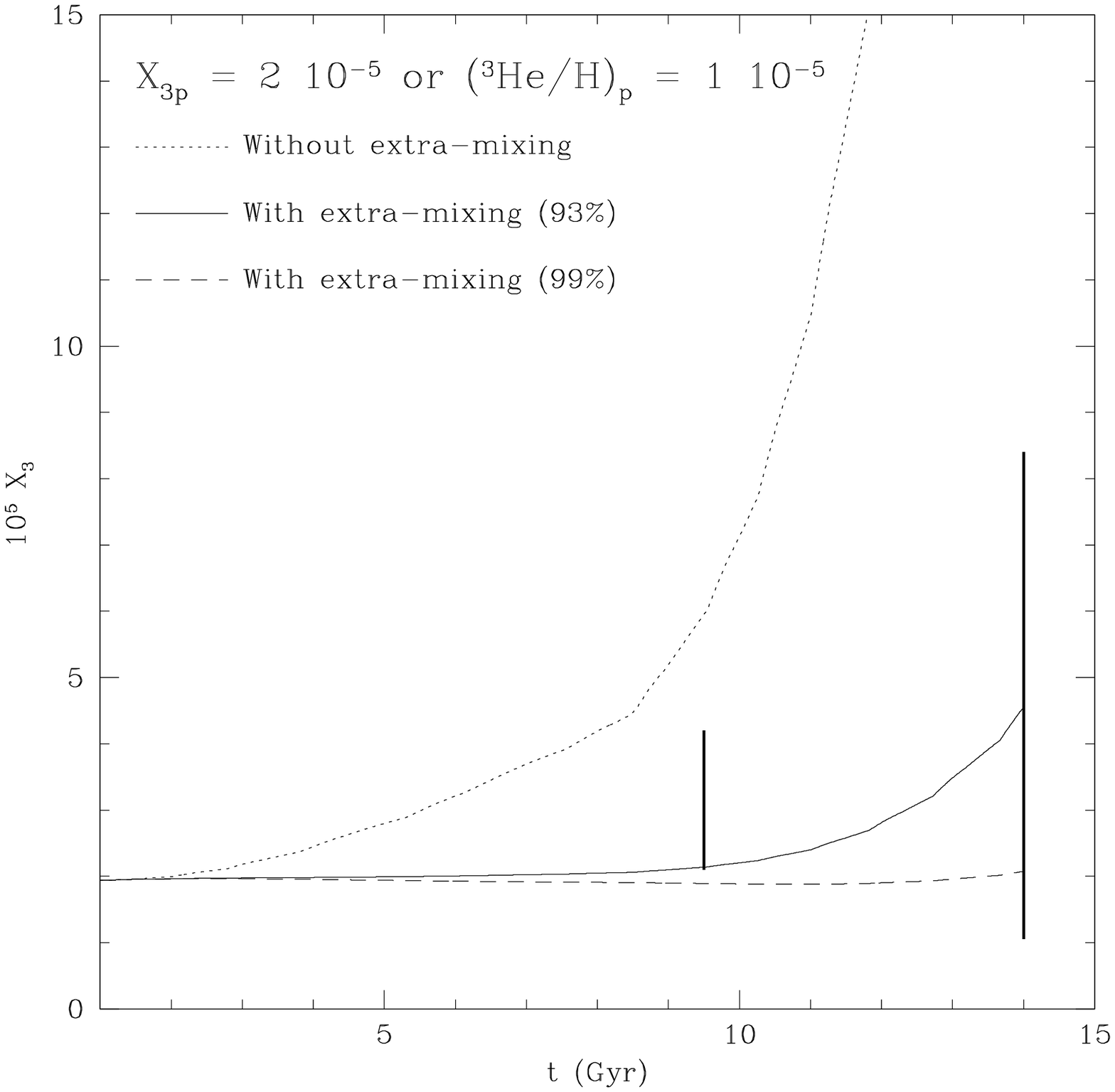,width=6.5cm,height=6cm}}
\caption{a) D and b) $^3$He evolution (by mass) as predicted by the present model. The bars at 9.5 Gyr (4.5 Gyrs ago) 
and 14 Gyr (age of the Galaxy) represent the value of solar (Geiss \& Gloeckler 1998 - 2$\sigma$) and ISM (Linsky 1998 - 2$\sigma$) measured abundances respectively. In the 
present model we are assuming a halo formation timescale of 1 Gyr.}
\end{figure*}

\begin{figure*}
\centerline{\psfig{figure=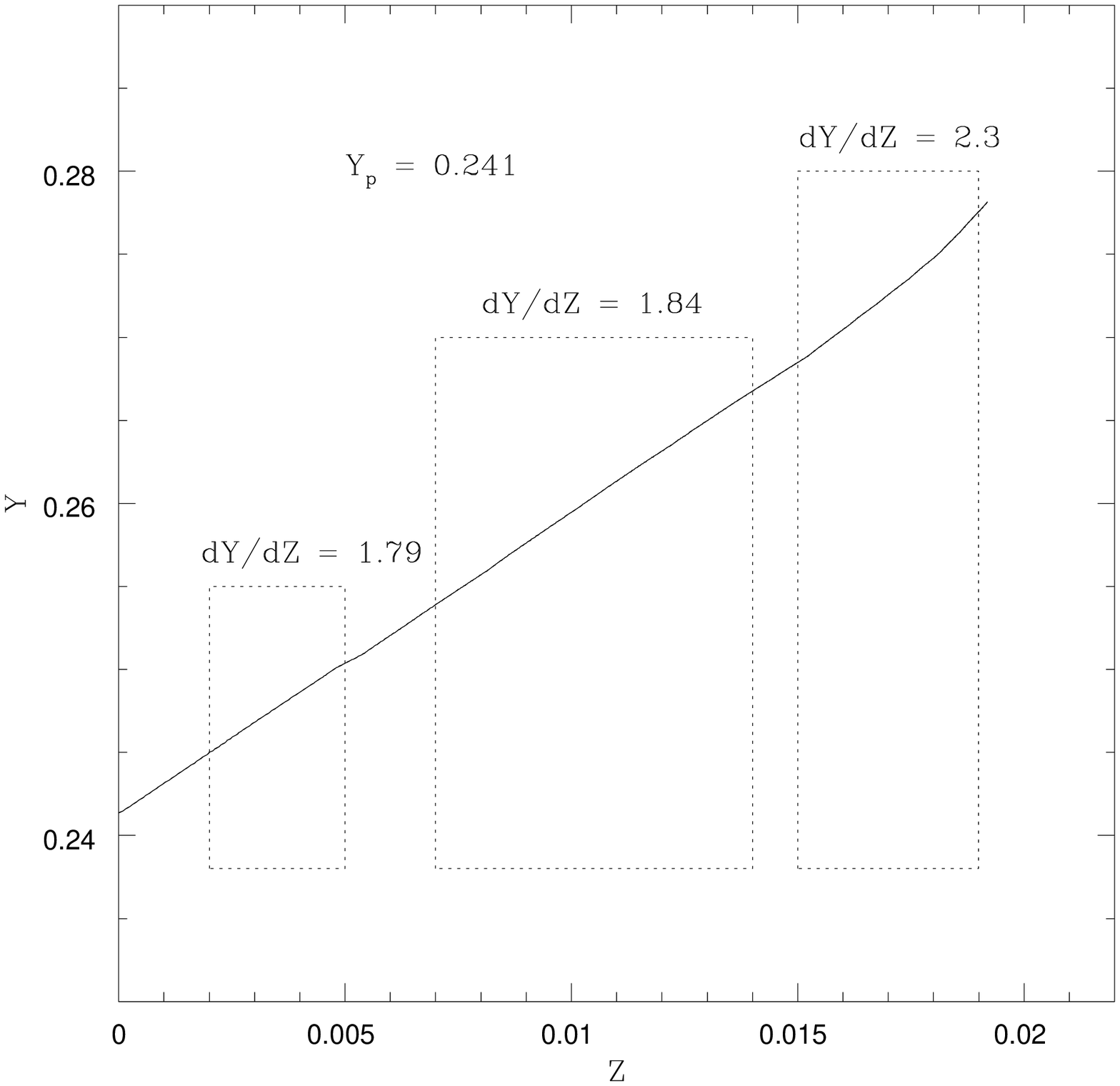,width=6.5cm,height=5.5cm}\psfig{figure=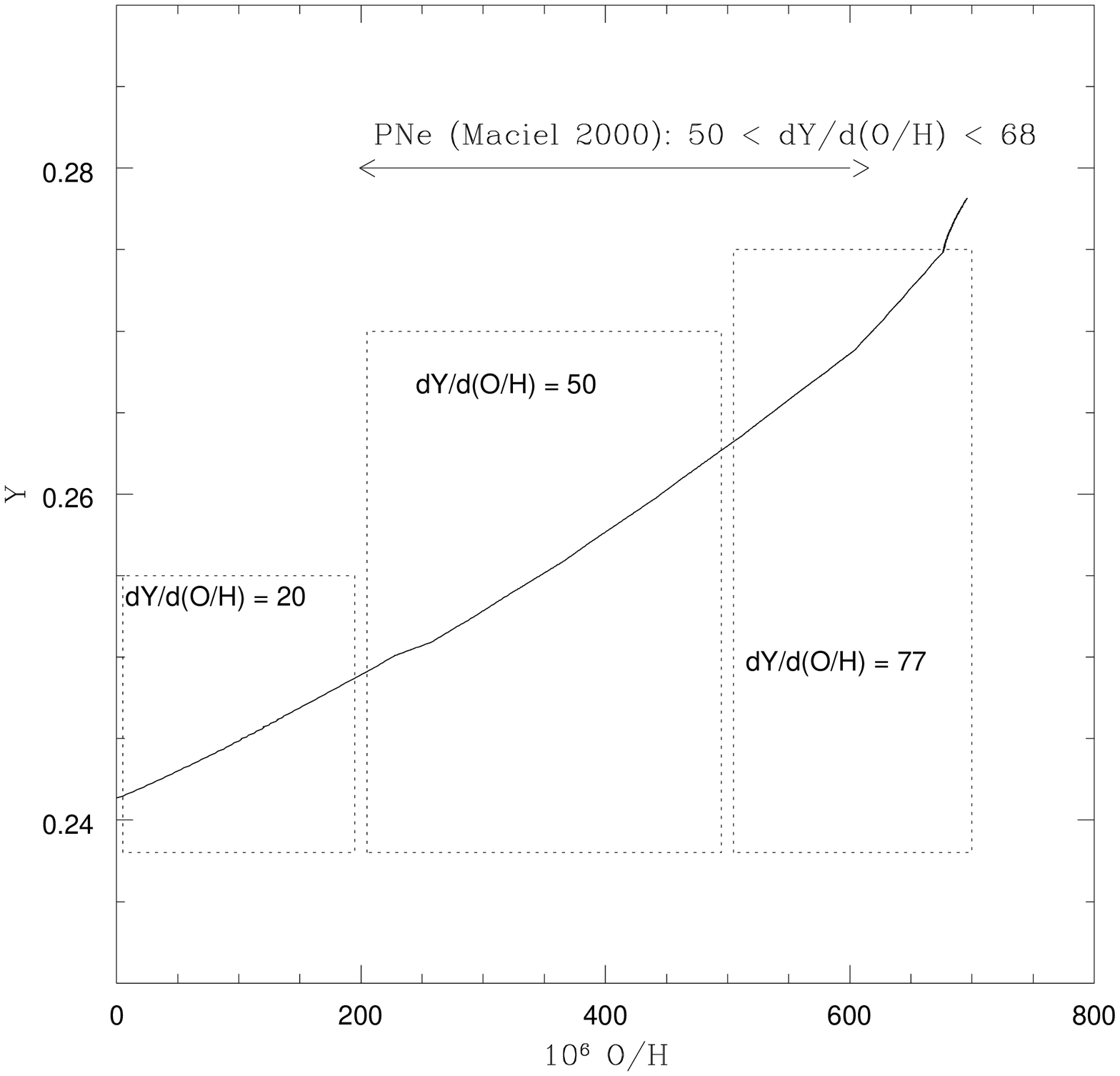,width=6.5cm,height=5.5cm}}
\caption{Helium by mass, Y, as a function of a) global metallicity, Z and b) O/H. The observed value obtained from 
planetary nebulae in the 200-600 10$^6$ O/H range is also shown.}
\end{figure*}

\subsection{The Galactic Disk} 

\begin{figure*}
\centerline{\psfig{figure=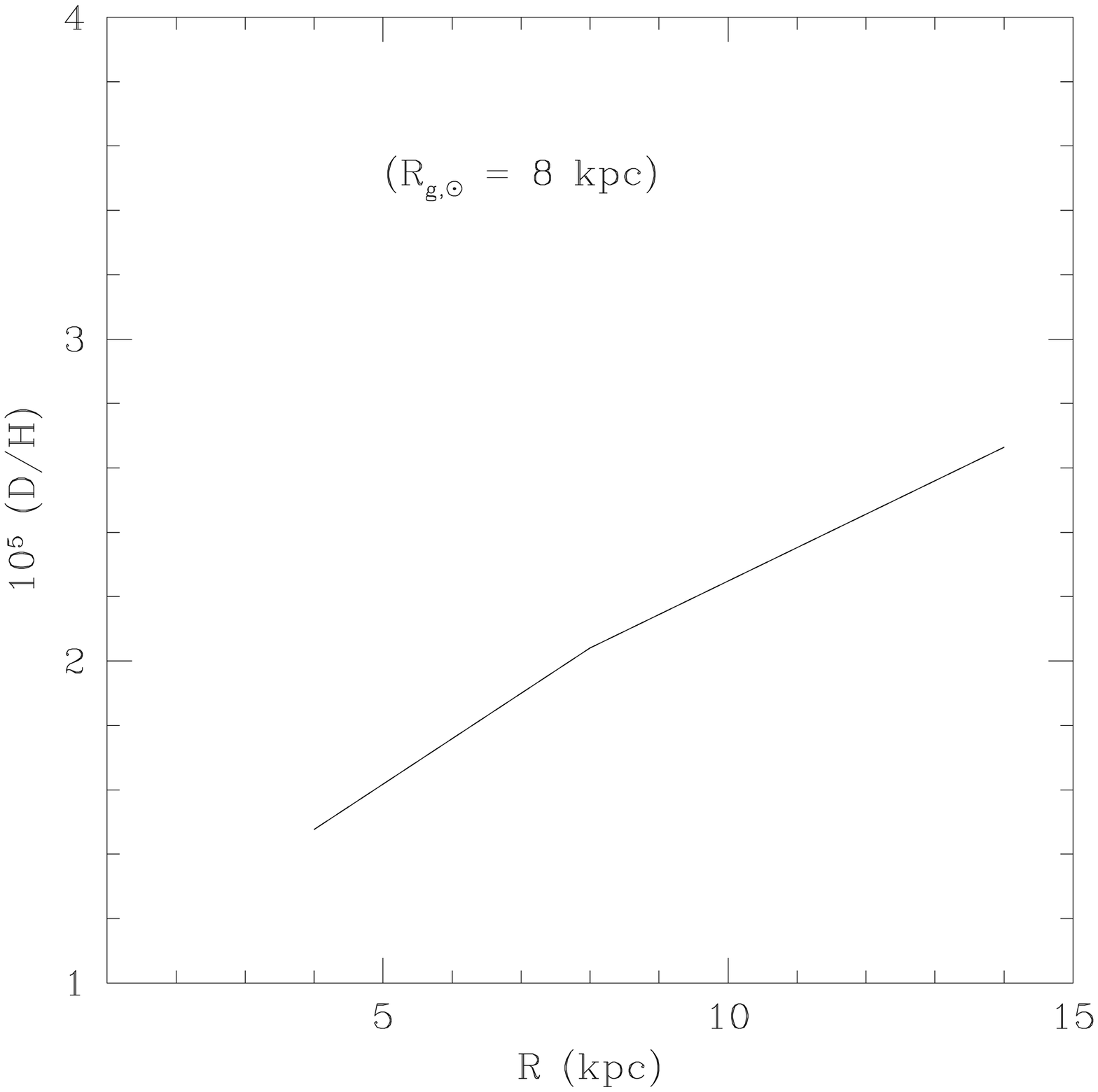,width=6.5cm,height=5.5cm}
\psfig{figure=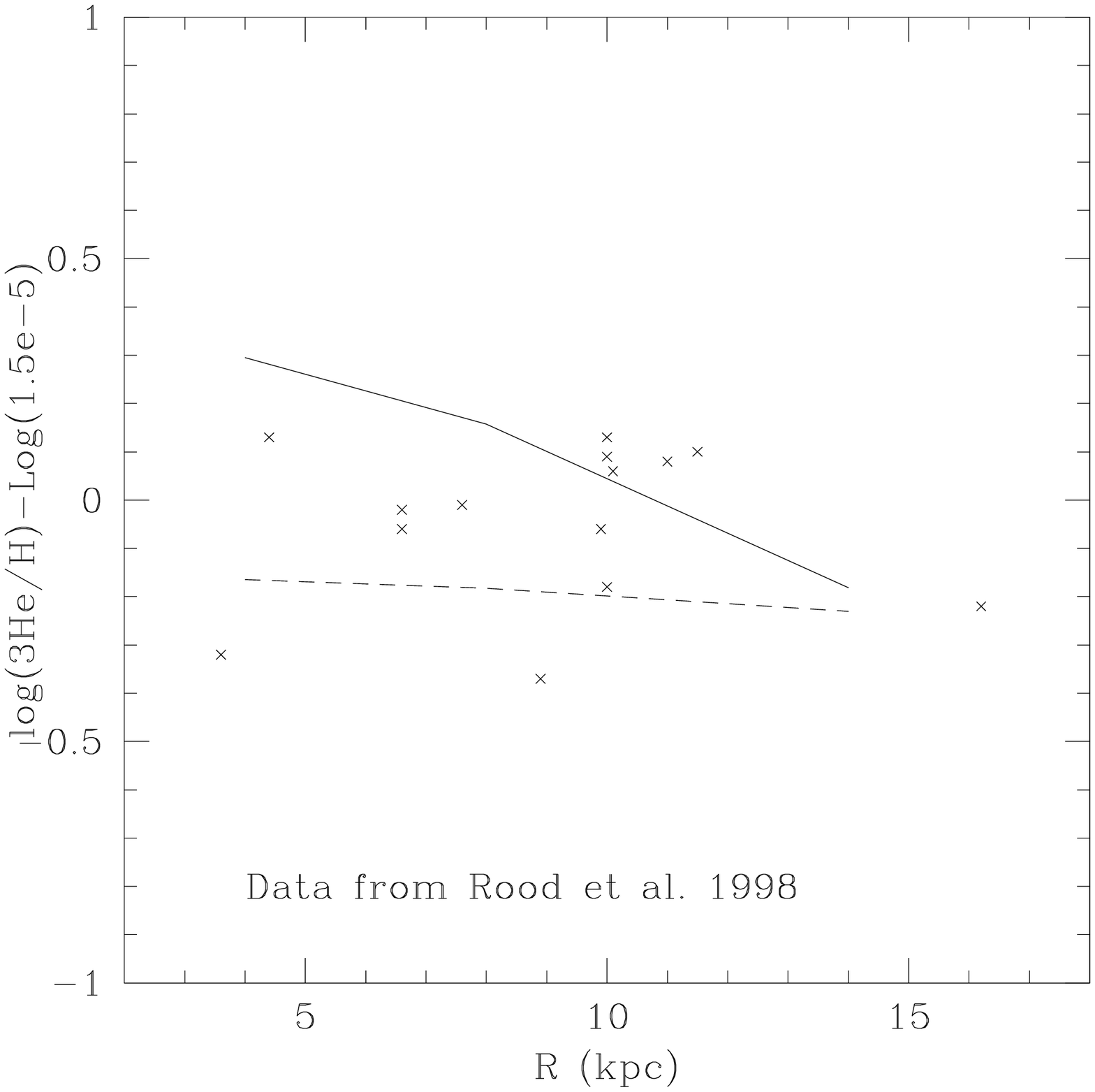,width=6.5cm,height=5.5cm}}
\caption{a) Predicted D gradient; b) Predicted $^3$He gradient. The curves are labelled as in figure 2b }
\end{figure*}

\begin{figure}
\centerline{\psfig{figure=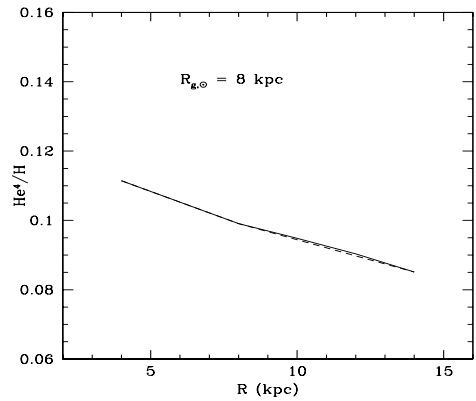,width=6.5cm,height=5.5cm}}
\caption{Predicted $^4$He/H abundance gradient}
\end{figure}

In this section we show our results for the predicted abundance distributions of D 
(Figure 4a),  $^3$He (Figure 4b)
and $^4$He (Figure 5).  The predicted gradient for D is positive and steep. This is due to the faster
evolution of the inner disk regions as compared with the outer parts (which are still in the process 
of formation thus having an almost primordial composition). 
In fact, of the various elements, D is probably the most sensitive to 
radial variations in the timescale of disk formation. A hope for the future is to have some D abundance
measurements in regions outside the solar vicinity. This would certainly represent a very important
constraint to the disk-formation mechanism ! 

In Figure 4b the predicted $^3$He abundance gradient is shown. It can be seen that the assumption that 99\% of
the low mass stars suffer extra-mixing leads to a flat gradient (dashed line). The solid line (93\% of low mass
stars destroy their $^3$He) is in marginal agreement with the data, but it is still acceptable. Again, in this case
we see that this gradient is sensitive to the adopted timescales of disk formation. In fact, in the inner
regions (older in the inside-out scenario) the contribution of low mass stars for the $^3$He enrichment of the 
ISM has been more important than in the outer regions, and a negative gradient is predicted.
More data on $^3$He abundance at different galactocentric distances are welcome and would be very important
to better constrain our models and the low-mass stellar nucleosynthesis as well.
Finally, in Figure 5 the $^4$He gradient is shown. The predicted gradient is $\simeq$ $-$0.003 dex/kpc over the 4-14kpc
galactocentric range. This value is in agreement with the results presented by Maciel (this meeting) based
on disk planetary nebulae.

\acknowledgments
C. C. thanks C. Charbonnel, M. Tosi, W. Maciel and B. Pagel for fruitful discussions during the meeting. The authors thank D. Romano who introduced in our code the new yields of van den Hoek \& Groenewegen (1997). 
C. C. acknowledges financial support by CNPq and FAPERJ (Brazil).

\end{document}